\documentclass[11pt]{article}
\usepackage[utf8]{inputenc}
\usepackage{amsmath, amssymb}
\usepackage{graphicx}
\usepackage{siunitx}
\usepackage{subcaption}
\usepackage{float}
\usepackage{hyperref}
\usepackage{authblk}
\usepackage{geometry}
\usepackage{acronym}
\acrodef{SiP}{silicon photonics}
\acrodef{PIC}{photonic integrated circuit}
\acrodef{FP}{Fabry-Perot}
\acrodef{CDM}{complementary dual-measurement}

\title{Fabry-Perot-Insensitive Edge Coupling for Robust PIC Characterization}

\author[1]{Sammy Noël Parisé}
\author[2]{Raphaël Dubé-Demers}
\author[1]{David Turgeon}
\author[1]{Alireza Geravand}
\author[1]{Sophie LaRochelle}
\author[1]{Wei Shi\thanks{Corresponding author: \texttt{wei.shi@gel.ulaval.ca}}}

\affil[1]{Department of Electrical and Computer Engineering, Université Laval, Québec, QC, Canada}
\affil[2]{EXFO Inc., Québec, QC, Canada}

\date{\small\textit{This manuscript is currently under review for publication in Photonics Technology Letters (IEEE).}}

\begin{document}
\maketitle

\section{Abstract}

In this work, we investigate the impact of \ac{FP} cavities formed between a fiber arrays and a silicon chips during edge-coupled testing of \acp{PIC}. Our results show that subwavelength variations in the distance between the fiber array and the device under test induce coupling efficiency modulations of several tenths of a decibel—challenging current alignment precision and undermining measurement repeatability in wafer-level probing. To address these FP-induced fluctuations, we introduce a complementary dual-measurement technique that shifts the chip–fiber separation by a quarter wavelength, thereby generating phase-opposite modulation artifacts that cancel upon averaging. This approach allows a higher tolerance in probe-to-PIC positionning by substantially eliminating the \ac{FP} modulation. We also demonstrate how it can be leveraged to enhance repeatability and accurate positionning during alignment, reducing the insertion loss variability between multiple measurements from 0.34 to 0.04 dB. Such precision is essential for emerging applications like quantum photonics, silicon photonic sensors, and high-bandwidth optical interconnects where measurement accuracy directly impacts device performance characterization. Ultimately, our method offers a robust and accurate solution for high-throughput PIC characterization without requiring index-matching gel, which is impractical for wafer-level testing.


\section{Introduction}

The rapid evolution of \acp{PIC} has been paralleled by an increasing demand for highly accurate and reliable characterization methods \cite{vermeulen_optical_2018, marchetti_coupling_2019}. Wafer-level testing has become mandatory for high-volume manufacturing in modern optical communication systems. With the recent advancement of manufacturable platforms for photonic quantum computing \cite{alexander_manufacturable_2025}, the precision requirements for these testing methodologies have reached unprecedented levels.
 While both surface coupling and edge coupling are widely employed, edge coupling has gained significant traction in high-performance and industrial applications due to its inherently lower insertion losses and reduced sensitivity to polarization and wavelength variations. However, this approach presents significant challenges in high-throughput testing environments where measurement accuracy is critical. Inaccuracies in these measurements can lead to discrepancies between pre- and post-packaging performance, ultimately undermining the reliability of the PICs in practical applications.

A critical yet often overlooked source of measurement inaccuracy in edge-coupled configurations is the formation of \ac{FP} cavities between the fiber array and the silicon chip. These cavities introduce resonant interference patterns that significantly affect insertion loss measurements. What makes this particularly challenging for wafer-level testing is that even subwavelength variations in the probe-to-chip distance—on the order of nanometers—can induce measurable fluctuations in coupling efficiency (up to 0.34 dB in our measurements). Traditional calibration approaches that rely on subtracting a reference measurement assume perfectly consistent coupling conditions across all measurements. However, achieving nanometer-precision positioning control between a fiber array and multiple devices across a wafer is impractical in industrial testing environments.
When testing individual dies, an index-matching gel is typically used to minimize this effect \cite{zhu_ultrabroadband_2016}. However, this approach becomes impractical when scaling to wafer or multi-die level testing, where the application and removal of gel would significantly complicate and slow the measurement process.

To address this challenge, we present a \ac{CDM} technique that not only reveals the presence of \ac{FP}-induced modulation errors in our experimental data but also proves highly effective in compensating for them through systematic phase manipulation. By performing two measurements at closely spaced, yet distinct, probe-to-PIC distances—specifically shifted by a quarter wavelength—we create datasets in which the \acp{FP} modulations are phase-opposed. When combined, these complementary modulations largely cancel each other, yielding a measurement of the transmission less sensitive to positionning variation. Our results, obtained by sweeping a tunable laser from 1505 nm to 1675 nm, confirm that this \ac{CDM} technique significantly mitigates insertion loss variability. We also studied the impact of the cavity on the alignment process in the $x$-$y$ plane. We demonstrate how the proposed method increases repeatability in both the insertion loss measurement and the positionning of the fiber array relative to the chip.

In the following sections, we detail the theoretical foundations, experimental setup, and empirical validation of our dual-measurement technique, underscoring its potential to improve the accuracy and reliability of PIC characterization in practical, high-throughput environments.

\section{Theory}

\begin{figure}[!h]
    \centering
    \includegraphics[width=0.75\linewidth]{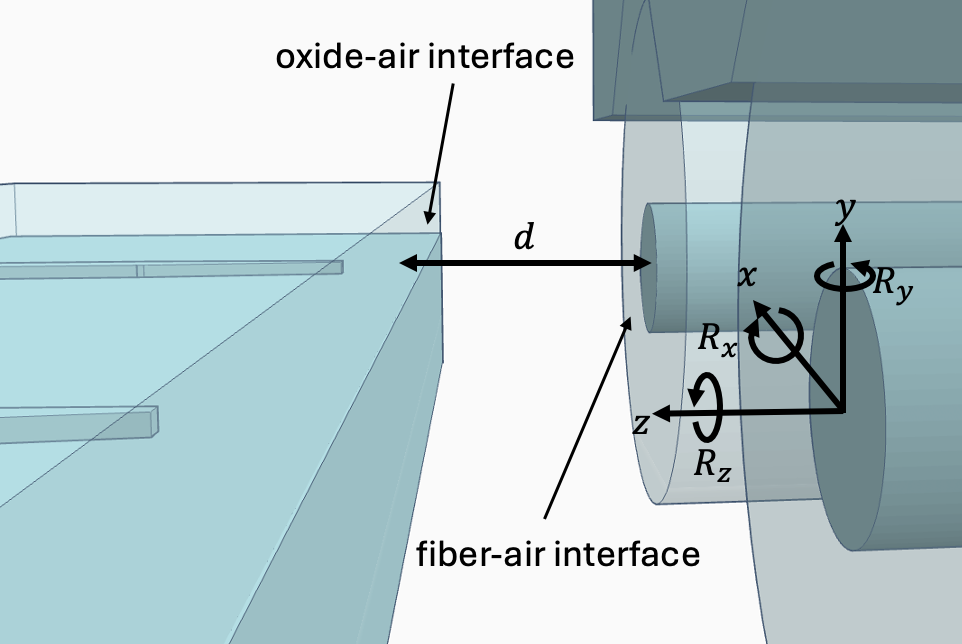}
    \caption{\ac{FP} cavity between the die and the optical fiber}
    \label{fig:fp}
\end{figure}
As illustrated on Fig \ref{fig:fp}, a \ac{FP} cavity is formed by two partially reflective surfaces—the fiber-air interface and the oxide-aire  interface—where multiple internal reflections produce resonant interference effects that depend on both wavelength and cavity length. These effects modulate transmitted light intensity, creating periodic enhancements or suppressions depending on the phase relationship between reflections \cite{vaughan_fabry-perot_2017}. In edge-coupled test configurations, the intrinsic circuit response is typically extracted by subtracting a loopback measurement from the measured response. However, both measurements are affected by \ac{FP} cavities that generate interference patterns, introducing systematic, position-sensitive fluctuations in insertion loss (IL) measurements. These compounding variations create significant uncertainty in the final calculated circuit response, undermining measurement reliability.
This position-dependent uncertainty—which we quantify as $\sigma_{FP}$ — presents a fundamental challenge in wafer-level testing. While modern positioning systems excel at precise relative movements, achieving consistent absolute positioning with subwavelength precision across multiple measurements remains impractical in production environments. Our \ac{CDM} technique transforms this challenging absolute positioning problem into a more tractable relative positioning solution by leveraging a controlled quarter-wavelength offset between measurements.

The transmission $T$ through a single \ac{FP} cavity formed of parallel interfaces is expressed as
\begin{equation}
  T = \frac{T_1\,T_2}{\,1 + R_1\,R_2 - 2\sqrt{R_1\,R_2}\cos\delta},
  \label{eq:Te}
\end{equation}
with $T_1$ and $T_2$ denote the transmission coefficients and $R_1$ and $R_2$ the reflection coefficients at the interfaces. The phase difference $\delta$ between successive reflections is defined as
\begin{equation}
  \delta = \frac{4 \pi n d}{\lambda},
\end{equation}
where $n$ is the refractive index of the medium ($\approx1$ for air) between the interfaces, $d$ is the cavity length, and $\lambda$ is the wavelength.

To assess the position-dependent uncertainty $\sigma_{FP}$ introduced by the \ac{FP} cavity, Eq.~\eqref{eq:Te} can be expanded in a Taylor series in powers of $\cos\delta$:
\begin{equation}
  T \approx \frac{T_1\,T_2}{1+R_1\,R_2}\left[ 1 + \frac{2\sqrt{R_1\,R_2}}{1+R_1\,R_2}\cos\delta + \cdots \right].
\end{equation}

At a silica--air interface ($n_{SiO_2}\approx 1.44$, $n_{air}\approx 1$) as found in fiber facets and edge couplers, the normal incidence reflection coefficient is
\begin{equation}
  R = \left|\frac{n_1 - n_2}{n_1 + n_2}\right|^2 \approx 0.032,
\end{equation}
yielding $T_1 \approx T_2 \approx 0.968$. Substituting these values into the expansion, we obtain
\begin{equation}
  T \approx 0.934\left(1 + 0.064\,\cos\delta + 0.004\,\cos^2\delta + \cdots\right).
\end{equation}
The uncertainty $\sigma_{FP}$ is dominated by the first-order term ($\cos\delta $) corresponding to a periodic sinusoidal fluctuation of up to approximately 0.29 dB, which is significant for precision measurements. The second-order term ($\cos^2\delta $) contributes only about 0.018 dB, making it negligible for most practical purposes. Since the dominant contributor to $\sigma_{FP}$ is the first-order term, our mitigation strategy focuses on eliminating this component.

To mitigate it, a \ac{CDM} technique is employed. By introducing a phase shift of $\delta' = \delta \pm \pi$, one obtains $\cos\delta' = -\cos\delta$. Since the only adjustable parameter is the distance between the interfaces, this phase shift is implemented by modifying the cavity length by translating the fiber array:
\begin{equation}
  d' = d \pm \frac{\lambda_0}{4n},
\end{equation}
where $\lambda_0$ is the design wavelength. With two measurements corresponding to cavity lengths $d$ and $d'$, the transmission coefficients are approximately
\begin{align}
  T &\approx 0.93\left(1 + 0.064\,\cos\delta + 0.004\,\cos^2\delta + \cdots\right), \\
  T' &\approx 0.93\left(1 - 0.064\,\cos\delta + 0.004\,\cos^2\delta + \cdots\right).
\end{align}
Averaging these measurements,
\begin{equation}
  T_{\mathrm{mean}} \approx \frac{T + T'}{2} \approx 0.93,
\end{equation}
effectively cancels the first-order error term, reducing $\sigma_{FP}$ from 0.17 dB to 0.04 dB. During a wavelength sweep, a residual error remains due to the wavelength-specific optimization of the phase shift. In general, the residual error is expressed as
\begin{equation}
  \Delta T(\lambda) = 0.064\,\cos\delta(\lambda) + 0.064\,\cos\delta'(\lambda),
\end{equation}
where
\begin{align}
  \delta(\lambda) = \frac{4\pi d}{\lambda}, \quad 
  \delta'(\lambda) = \frac{4\pi }{\lambda} \left( d + \frac{\lambda_0}{\lambda}\right).
\end{align}
This residual error is minimized when $\lambda_0\approx \lambda$. For example, with $\lambda = \SI{1.505}{\micro\meter}$ and $\lambda = \SI{1.675}{\micro\meter}$, and choosing $\lambda_0 = \SI{1.6}{\micro\meter}$ (values used in our experiments), the maximum errors $\sigma_{FP}(\lambda)$ are approximately \SI{0.05}{dB} and \SI{0.04}{dB}, respectively.

\section{Experimental Setup}
\begin{figure}[t]
\centering
\includegraphics[width=4.5in]{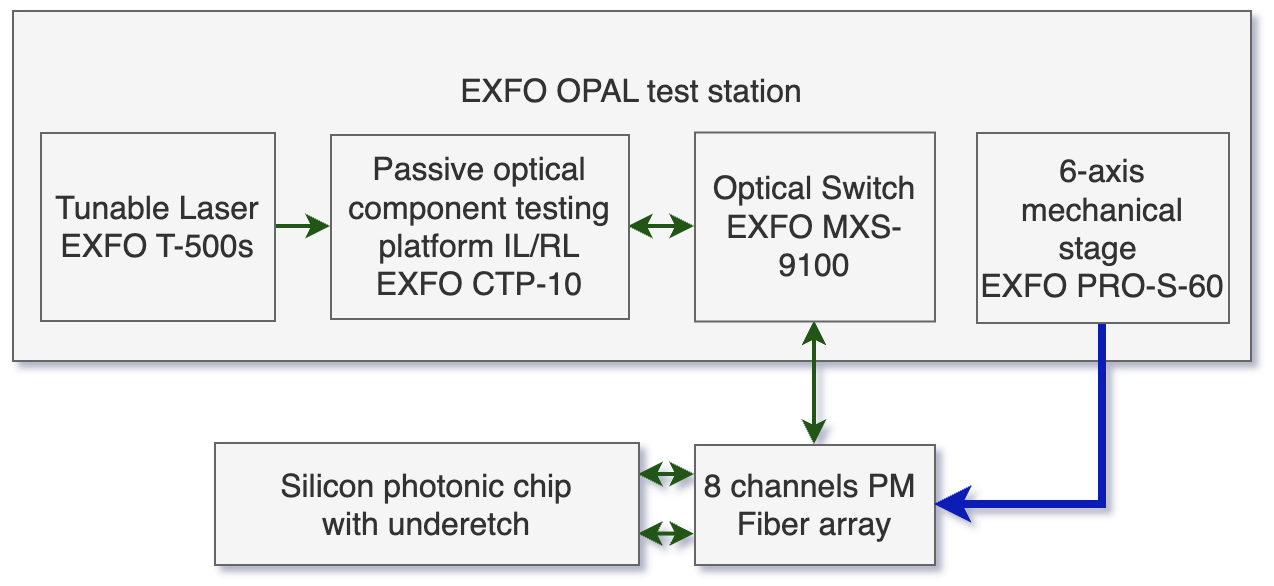}
\caption{Experimental setup for the edge coupling measurements. A sweeping tunable laser is directed through an IL/RL module, optical switch, and fiber array to the silicon photonic chip. The optical return loss is monitored by a photodetector inside the IL/RL module.}
\label{setup.png}
\end{figure}

Figure \ref{setup.png} illustrates the experimental setup used in our study. A tunable laser, sweeping wavelengths from 1505 to 1675 nm, served as the light source and was routed through an IL/RL module (Insertion Loss \& Return Loss) with an integrated circulator and photodetector to monitor the optical return loss. The conditioned output was then directed into an optical switch, which fed the signal into a fiber array cleaved at 0°. The fiber array was mounted on a high-precision 6-degrees of freedom mechanical stage with a minimal increment of 3 nm, allowing for controlled sub-wavelength adjustments of the probe-to-PIC gap that are critical for our phase-compensation technique. The light was subsequently coupled into a loopback configuration with suspended edge couplers \cite{jia_efficient_2018}. The EXFO PILOT$^{\text{TM}}$ software was used to control the test station, ensuring accurate and repeatable measurements throughout the experimental process.

\vspace{-5pt}
\section{Experimental Results and Analysis}

\begin{figure}[h]
    \centering
    \begin{subfigure}[b]{0.45\textwidth}
        \includegraphics[width=\textwidth]{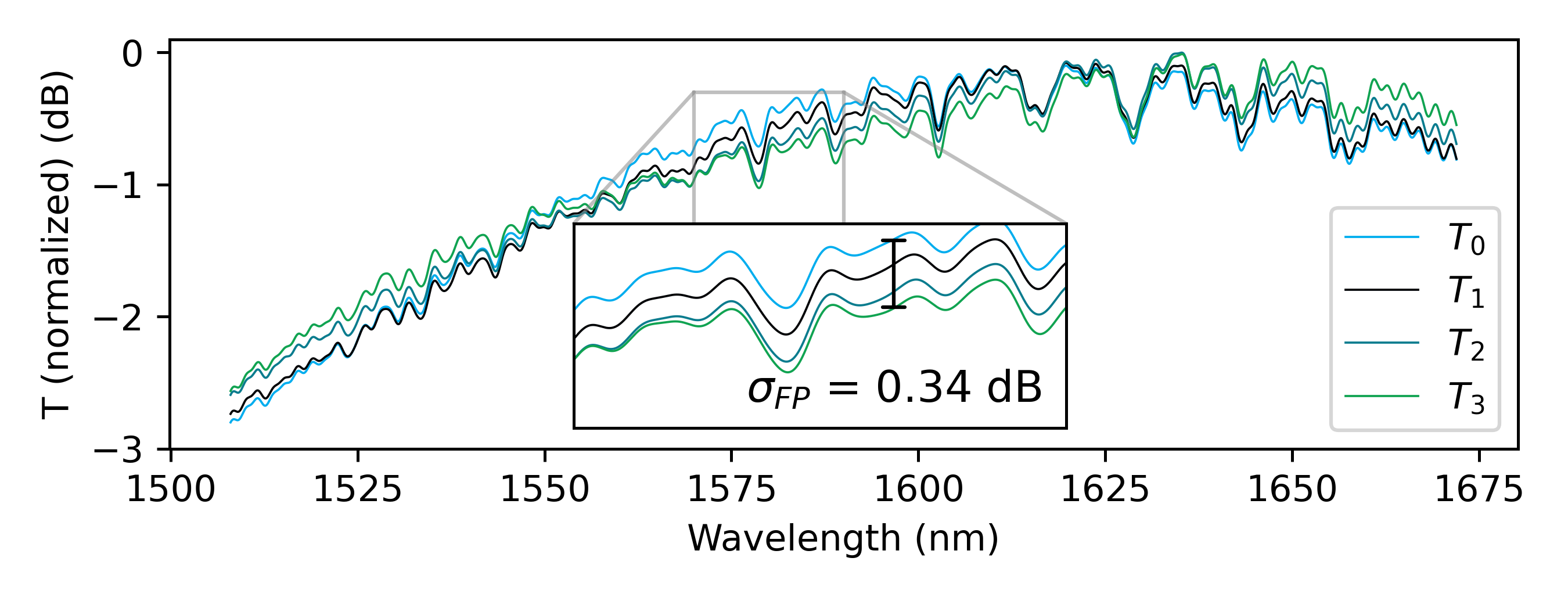}
        \caption{Single measurement}
        \label{fig:single}
    \end{subfigure}
    \begin{subfigure}[b]{0.45\textwidth}
        \includegraphics[width=\textwidth]{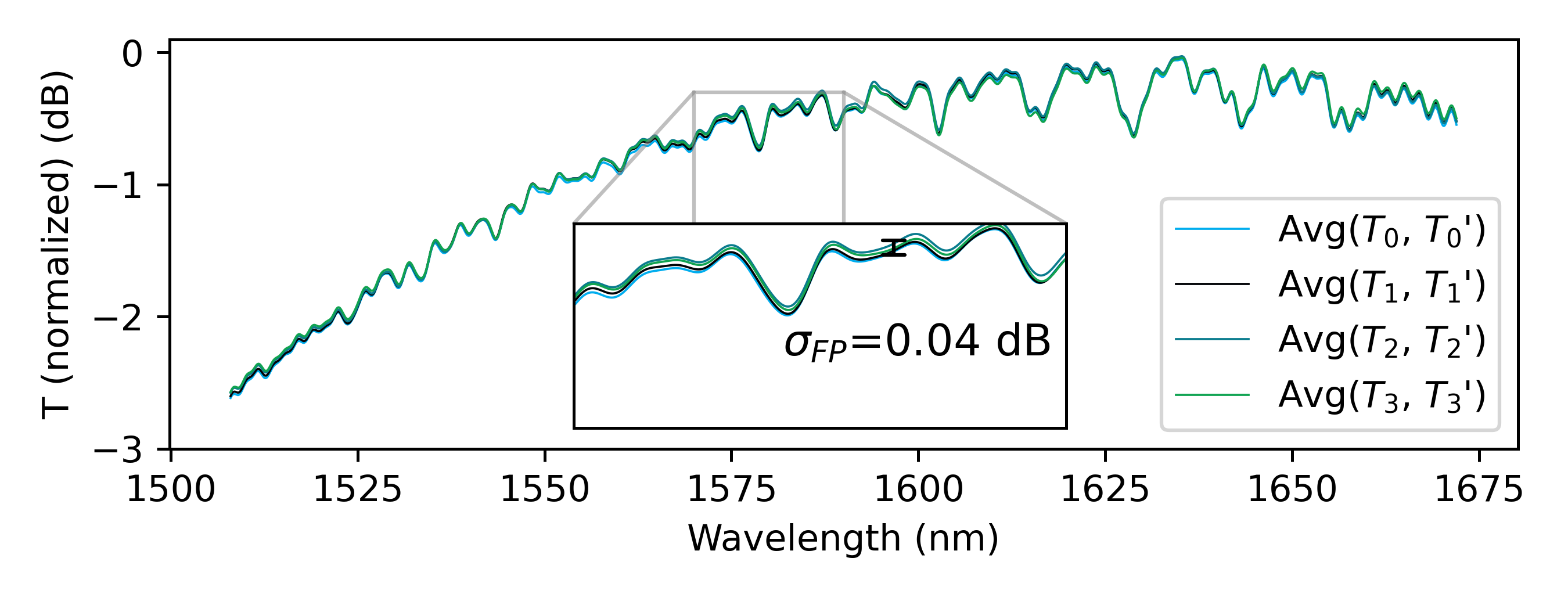}
        \caption{Complementary dual-measurements}
        \label{fig:dual}
    \end{subfigure}
    \begin{subfigure}[b]{0.8\textwidth}
        \includegraphics[width=\textwidth]{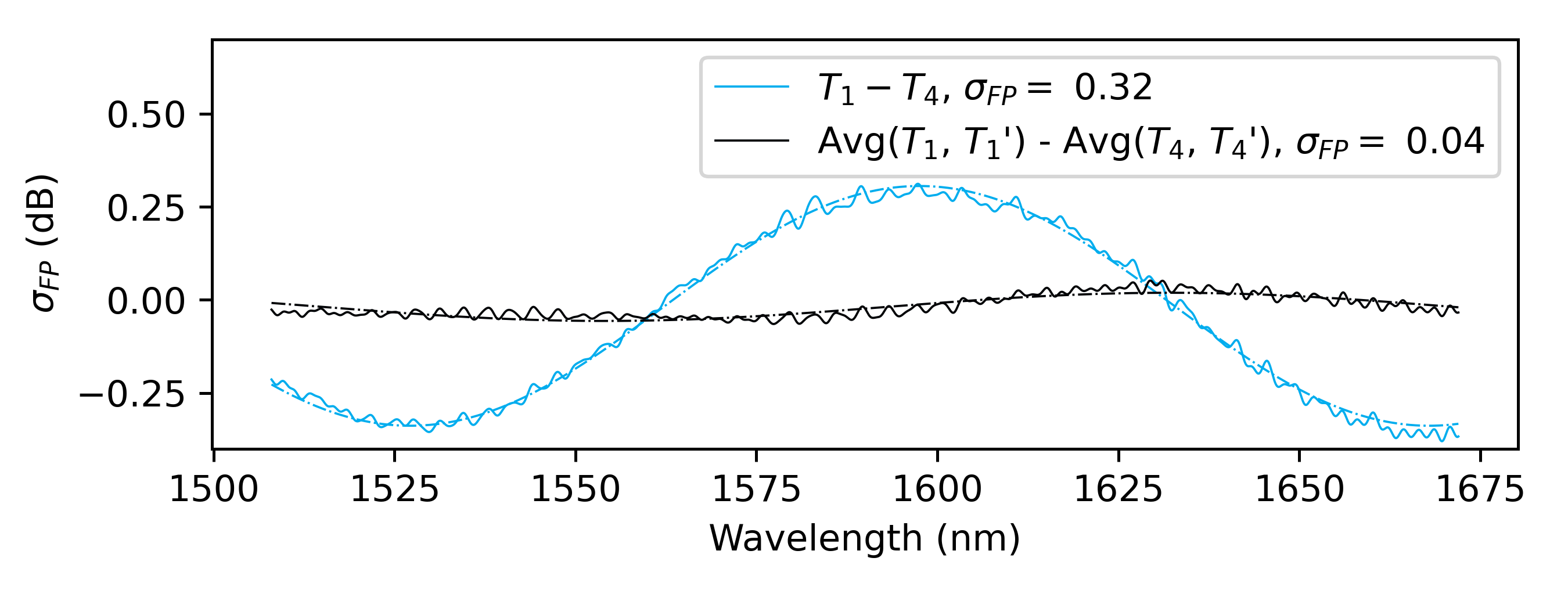}
        \caption{Difference between measurements}
        \label{fig:difference}
    \end{subfigure}
    \caption{Comparison of spectra acquired by a single sweep and \ac{CDM}. Fig \ref{fig:single} show the transmission spectra of the same edge coupler loopbacks where the chip-fiber distance was shifted by 0.1 $\mu$m between each measurements. Fig \ref{fig:dual} shows the same spectra where two measurements acquired with a 0.4 $\mu$m shift in the fiber-chip distance were averaged together, cancelling the \ac{FP} modulation. Fig \ref{fig:difference}
    shows a comparison in the fluctuation between two measurement at slightly difference distance using single and \ac{CDM} measurement.}
    \label{fig:overall}
\end{figure}

A serie of insertion loss (IL) measurements were performed by sweeping a tunable laser over a 170 nm range. Between each measurement, the fiber array was shifted by 100\,nm along the optical axis to modify the \ac{FP} cavity length. Figure~\ref{fig:single} presents four smoothed IL spectra obtained under these conditions. Although a Gaussian filter was applied to reduce high-frequency variations, significant noise remained in the raw data, which we primarily attribute to polarization-dependent fluctuations in the automated optical switch. Despite this residual noise, the modulation induced by the \acp{FP} cavity can be seen in the single measurements, with spectral variations $\sigma_{FP}$ in the insertion loss reaching up to 0.34\,dB as it can be observed on fig~\ref{fig:single}

Fig. \ref{fig:dual} displays the \ac{CDM}, derived by combining two complementary measurements separated by 400 nm. In contrast to the unpaired data—which exhibit a broad variation band—the \ac{CDM} show a considerably narrower spread. As explained in the theory section, this pairing effectively cancels the dominant \acp{FP}-induced ripple, resulting in a more stable IL measurement resulting in a maximal spectra variation $\sigma_{FP} $ of 0.04 dB. Fig \ref{fig:difference}
 illustrates how a variation in cavity gap between the PIC and the fiber array introduces a variation in the insertion loss across the spectrum. The \ac{CDM} approach reduces the \ac{FP}-related variation to 0.04 dB.

Fig. \ref{fig:dual} demonstrates the effectiveness of the \ac{CDM} technique, which combines two complementary measurements separated by a quarter wavelength. While the unpaired measurements exhibit significant spectral variation with pronounced ripple patterns, the \ac{CDM} approach shows a markedly reduced spread. This improvement occurs because, as detailed in the theory section, the pairing mathematically cancels the dominant \ac{FP}-induced ripple components. The result is a more stable insertion loss measurement across the spectrum, with the maximal spectral variation ($\sigma_{FP}$) reduced to only 0.04 dB. Fig. \ref{fig:difference} highlights the characteristic sinusoidal nature of these \ac{FP}-induced artifacts when substracting opposed phase measurements.

We further investigated the practical impact of \ac{FP} cavity interference on focal-plane alignment procedures, which is crucial for automated testing systems where alignment reliability directly affects throughput and measurement quality. Two measurement schemes were evaluated: a conventional single-measurement approach and a the \ac{CDM} technique. For each, the coupling efficiency was optimized by adjusting the fiber array position using a Nelder–Mead algorithm over 800 iterations. During each optimization run, both the coupling power (in dB) and the position in the $x$-$y$ plane were recorded.  

Prior to analysis, the acquired data were preprocessed to ensure consistency. Outliers were removed by retaining only those data points within \(\pm2\sigma\) of the mean for each variable, and the analysis was restricted to the first 700 points to maintain uniformity between both dataset. In order to confirm that the observed error reduction was not simply due to averaging, a control analysis was performed in which each single measurement was averaged with the subsequent one, mimicking the averaging effect of the \ac{CDM}.

\begin{figure}[H]
  \centering
  \includegraphics[width=0.96\textwidth]{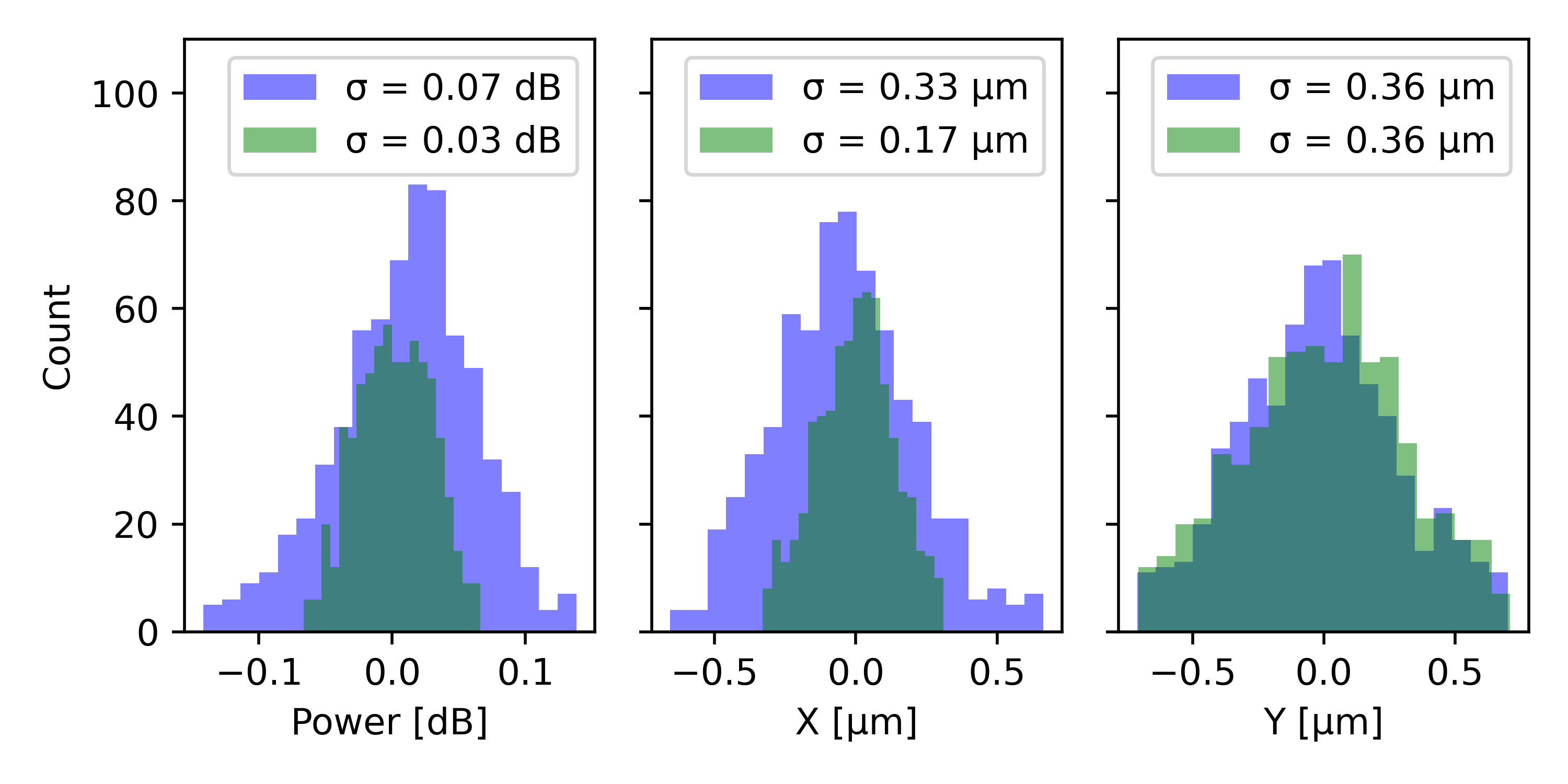}
  \caption{Histograms showing the distribution of measurement errors for single and paired alignment approaches. Left: Coupling efficiency (power) variations in dB, demonstrating the repeatability of optical power coupling. Middle: X-axis positional variations in µm, showing horizontal alignment precision. Right: Y-axis positional variations in µm, indicating vertical alignment repeatability. In all three histograms, blue distributions represent single measurements, while green distributions represent the Complementary Dual Measurement (\ac{CDM}) technique. The standard deviation ($\sigma_z$) values quantify the measurement repeatability for each method and parameter.}
  \label{fig:repeatability}
\end{figure}

The resulting distributions of residual errors are shown as histograms in Fig.~\ref{fig:repeatability}. The histograms reveal that the \ac{CDM} approach produces markedly narrower distributions compared to the single-measurement method. This is evident from the lower standard deviations observed for both the coupling power and the lateral (X) alignment positions observed of Fig. \ref{fig:repeatability}. The reduced variability confirms that the use of complementary data effectively suppresses the \acp{FP}-induced noise, leading to more robust and repeatable fiber array alignment. 

It is important to note that while the Nelder–Mead algorithm is designed to search for a local optimum in a smooth landscape. The strong position dependence of the FP insertion loss modulation appears to create a complex landscape with multiple local extrema. We hypothesize that these local minima (and potentially local maxima) can mislead the algorithm, causing it to converge to suboptimal solutions rather than the true global optimum. These findings not only confirm our theoretical predictions regarding \ac{FP} error cancellation but also underscore the significant potential of this method for wafer-level and high-volume PIC testing applications where precise, repeatable measurements are essential for quality control and device performance characterization. The technique can be implemented on existing test platforms without additional hardware, making it immediately applicable to industrial testing environments.

Interestingly, although the double-measurement technique significantly reduced errors in both the coupling power and the X position, a similar improvement was not observed along the Y axis. We suspect that this discrepancy might be due to a slight angular misalignment in the experimental setup, leading to a non-normal incidence in the $x$-$y$ plane (refer to fig \ref{fig:fp}  for coordinate system). Such a condition could result in a less smooth irradiance profile along the Y direction \cite{abu-safia_transmission_1994}, thereby increasing the probability that the Nelder–Mead algorithm converges to local extrema instead of the global optimum. While this remains a hypothesis, these observations suggest that further investigation into the influence of angular misalignment on the optimization process is warranted.

These findings validate the  \ac{CDM} technique as a viable means to enhance alignment performance in integrated photonic systems, while also indicating potential avenues for future research to address the observed asymmetry in error reduction.

\vspace{-5pt}
\section{Conclusion}

In this study, we introduced a \ac{CDM} technique to mitigate the systematic errors induced by \ac{FP} cavities in insertion loss and alignment measurements for edge-coupled photonic integrated circuits. This novel approach addresses a critical yet often overlooked challenge in wafer-level testing, where even subwavelength variations in probe-to-chip distance can significantly impact measurement accuracy and repeatability.
By carefully adjusting the fiber array position by a quarter wavelength and combining measurements with complementary FP-induced ripples, we achieved a substantial reduction in measurement variability. The CDM technique effectively canceled the dominant first-order error terms, dramatically improving measurement repeatability across the spectral range. This method also enhanced the repeatability of the alignment process in the $x$-$y$ plane, particularly for coupling efficiency and X-axis positioning.
These findings confirm our theoretical predictions regarding FP error cancellation and demonstrate that our approach can be implemented on existing test platforms without additional hardware, making it immediately applicable to industrial testing environments. As photonic integrated circuits continue to advance into increasingly demanding applications, the measurement precision enabled by our technique will play a crucial role in ensuring reliable device characterization and performance verification.

\section*{Acknowledgment}
We want to acknowledge CMC Microsystems, manager of the FABrIC project funded by the Government of Canada.

\vspace{-5pt}
\bibliography{references}
\bibliographystyle{plain}
\end{document}